%% file: main.tex
\documentclass[fleqn,usenatbib]{mnras}

\usepackage{newtxtext,newtxmath}

\usepackage[T1]{fontenc}
\usepackage{ae,aecompl}

\usepackage{graphicx}	
\usepackage{amsmath}	
\usepackage{amssymb}	
\usepackage{hyperref}

\input{commands}

\def\bee{\begin{eqnarray}}
\def\ene{\end{eqnarray}}

\title[Filament lensing with SDSS BOSS and HSC]{Weak Lensing Measurement of Filamentary Structure with the SDSS BOSS and Subaru Hyper Suprime-Cam Data}

\author[H. Kondo et al.]{
Hiroto Kondo,$^{1}$\thanks{E-mail: kondo.hiroto@h.mbox.nagoya-u.ac.jp}
Hironao Miyatake,$^{2,1,3}$
Masato Shirasaki,$^4$
Naoshi Sugiyama,$^{1,3}$
\newauthor
Atsushi J. Nishizawa$^{2,1}$
\\
$^{1}$Division of Particle and Astrophysical Science, Graduate School of Science, Nagoya University, Furo-cho, Nagoya 464-8602, Japan\\
$^{2}$Institute for Advanced Research, Nagoya University, Furo-cho, Nagoya 464-8601, Japan\\
$^{3}$Kavli Institute for the Physics and Mathematics of the Universe (Kavli IPMU, WPI), UTIAS, The University of Tokyo, Kashiwa, \\Chiba 277-8583, Japan\\
$^{4}$National Astronomical Observatory of Japan, Mitaka, Tokyo 181-8588, Japan
}

\date{Accepted XXX. Received YYY; in original form ZZZ}

\pubyear{2019}

\begin{document}
\label{firstpage}
\pagerange{\pageref{firstpage}--\pageref{lastpage}}
\maketitle

\begin{abstract}
We report the weak lensing measurement of filaments between Sloan Digital Sky Survey (SDSS) III/Baryon Oscillation Spectroscopic Survey (BOSS) CMASS galaxy pairs at $z\sim0.55$, using the Subaru Hyper Suprime-Cam (HSC) first-year galaxy shape catalogue. Despite of the small overlap of $140$~deg$^2$ between these surveys we detect the filament lensing signal at 3.9$\sigma$ significance, which is the highest signal-to-noise lensing measurement of filaments between galaxy-scale halos at this redshift range. We derive a theoretical prediction and covariance using mock catalogues based on full-sky ray-tracing simulations. We find that the intrinsic scatter of filament properties and the fluctuations in large scale structure along the line-of-sight are the primary component of the covariance and the intrinsic shape noise from source galaxies no longer limits our lensing measurement. This fact demonstrates the statistical power of the HSC survey due to its deep observations and high number density of source galaxies. Our result is consistent with the theoretical prediction and supports the ``thick'' filament model. As the HSC survey area increases, we will be able to study detailed filament properties such as the dark matter distributions and redshift evolution of filaments.
\end{abstract}




\section{Introduction}
\label{sec:introduction}
One of the key features of what the $\Lambda$ cold dark matter ($\Lambda$CDM) model predicts is that the large scale structure of the Universe, which is formed from small scales such as stars and galaxies to large scales such as galaxy groups and clusters. In this process, the CDM collapses into dark matter halos and then galaxies are formed within a dark matter halo as a result of gas accretion. After a dark matter halo evolves, it hosts more galaxies and then galaxy groups and clusters are formed. Such massive halos are connected to each other via filamentary structure, the so-called cosmic web. Such filamentary structures have been clearly seen in the three-dimensional galaxy distribution observed by galaxy redshift surveys such as the 2dF Galaxy Redshift Survey \citep[2dFGRS;][]{Colless:2001} and Sloan Digital Sky Survey \citep[SDSS;][]{York:2000}. Statistical measurement of the filamentary structure in the galaxy overdensity between massive halos has also been performed by stacking pairs of galaxy groups and clusters \citep[e.g.,][]{Zhang:2013} 

In addition to these measurements, it is of great importance to measure the dark matter filaments, which will be a more direct proof of the $\Lambda$CDM cosmology. Such a measurement can be made possible by weak gravitational lensing which manifests as a coherent distortion in observed galaxy shapes caused by foreground dark matter structures.

Several weak lensing measurements of a filament either between massive galaxy clusters or connected to a massive galaxy cluster are reported. \citet{Dietrich:2012} detected a dark matter filament in the super cluster system consisting of Abell 222 and Abell 223. \citet{Jauzac:2012} claimed the detection of a dark matter filament funnelling onto a core of massive galaxy cluster MACSJ0717.5+3745. \citet{Higuchi:2015} measured the average surface mass density of a dark matter filament between galaxy clusters at $z=0.55$, CL0015.9+1609 and RX J0018.3+1618.

Several attempts to make a statistical detection of dark matter filaments by stacking pairs of galaxies have been made. \citet{Clampitt:2016} reported the detection of dark matter filaments at 4.5$\sigma$ significance by stacking 135,000 Luminous Red Galaxy (LRG) pairs from SDSS with the typical redshift $z\sim0.25$, and claimed that their filament lensing signal is consistent with a ``thick'' filament model predicted from dark matter $N$-body simulations rather than a ``thin'' filament model by a string of halo along the line connecting the two LRGs. \citet{Epps:2017} pushed the pair redshift to $\langle z \rangle \sim 0.42$ using the SDSS-III/Baryon Oscillation Spectroscopic Survey \citep[BOSS;][]{Dawson:2013} LOWZ and CMASS galaxy sample, and claimed the detection at 5$\sigma$ significance. \citet{He:2018} claimed the filament lensing detection at 5$\sigma$ significance using the filament catalogue constructed from density ridges of CMASS galaxies \citep{Chen:2016} and weak lensing of cosmic microwave background (CMB) measured from the Planck satellite experiment \citep{Planck2013I, Planck2013XVII}.

In this paper, we report the weak lensing measurement of filaments between the SDSS-III/BOSS CMASS galaxy pairs using the Subaru Hyper Suprime-Cam \citep[HSC;][]{Miyazaki:2018, Komiyama:2018, Kawanomoto:2018, Furusawa:2018} Subaru Strategic Program \citep[SSP;][]{Aihara:2018a} first-year data \citep{Aihara:2018b}. Among the ongoing weak lensing surveys such as Dark Energy Survey \citep[DES;][]{DES:2016} and the Kilo-Degree Survey \citep[KiDS;][]{KiDS:2013}, the HSC survey provides the deepest images, which enables us to perform one of the highest signal-to-noise measurements of filament weak lensing. We also use the mock CMASS and source galaxy catalogue to estimate theoretical prediction and covariance.

This paper is organised as follows. In Section~\ref{sec:data}, we describe the details of the data set used for our measurement, including CMASS galaxy pair generation, HSC galaxy shape catalogue, and mock catalogue generation. In Section~\ref{sec:measurement}, we describe measurement methods such as filament lensing estimator and covariance. We then present the results of our filament lensing measurement in Section~\ref{sec:result}, and conclude in Section~\ref{sec:conclusions}. Throughout the paper, we assume the Planck 2015 cosmology \citep{Planck2015XIII}, i.e., $\Omega_{\rm m}=0.307$ with the flat $\Lambda$CDM, unless otherwise stated.

\section{Data}
\label{sec:data}
\subsection{CMASS Galaxy Pairs}
\label{sec:data_boss}
The goal of this paper is to measure the filamentary dark matter distribution between massive galaxies by stacking weak lensing signal around the galaxy pairs. To construct the galaxy pair catalogue, we use the SDSS-III/BOSS CMASS large-scale structure (LSS) galaxy catalogue \citep{Alam:2015} with the redshift cut $0.14 < z < 0.88$  Note that we do not impose any selections other than the redshift cut.

The spectroscopic redshift of the CMASS sample enables us to select galaxy pairs with the robust measurement of the line-of-sight distance. In contrast, if we rely on a photometric sample, the typical statistical uncertainty of a photometric redshift $\Delta z_{\rm p} \sim 0.1$, which corresponds to the comoving distance $\Delta \chi \sim 200~\mpch$ at the typical CMASS redshift, introduces more galaxies pairs without filaments and thus dilutes the filament lensing signal.

\begin{figure}
	\includegraphics[width=\columnwidth]{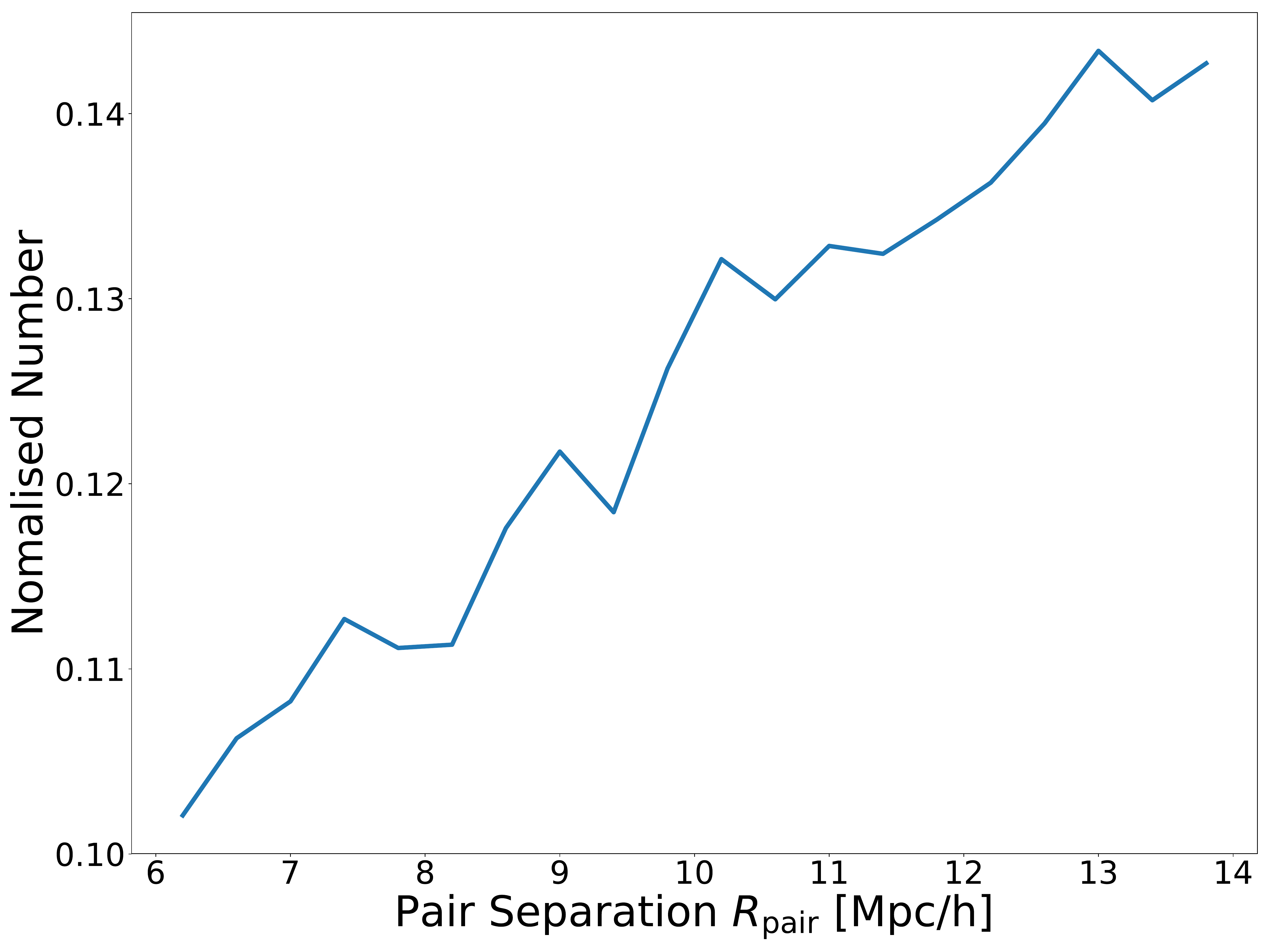}
    \caption{The distribution of physical transverse separation between CMASS galaxy pairs used for our analysis. The y-axis is normalised such that $\sum  \Delta R_{\rm pair} n(R_{\rm pair})=1$, where $n(R_{\rm pair})$ is the normalised number of pairs shown in this figure and $\Delta R_{\rm pair}$ is the bin width with our choice of $\Delta R_{\rm pair}=0.4$.}
    \label{fig:separation_distribution}
\end{figure}

\begin{figure}
	\includegraphics[width=\columnwidth]{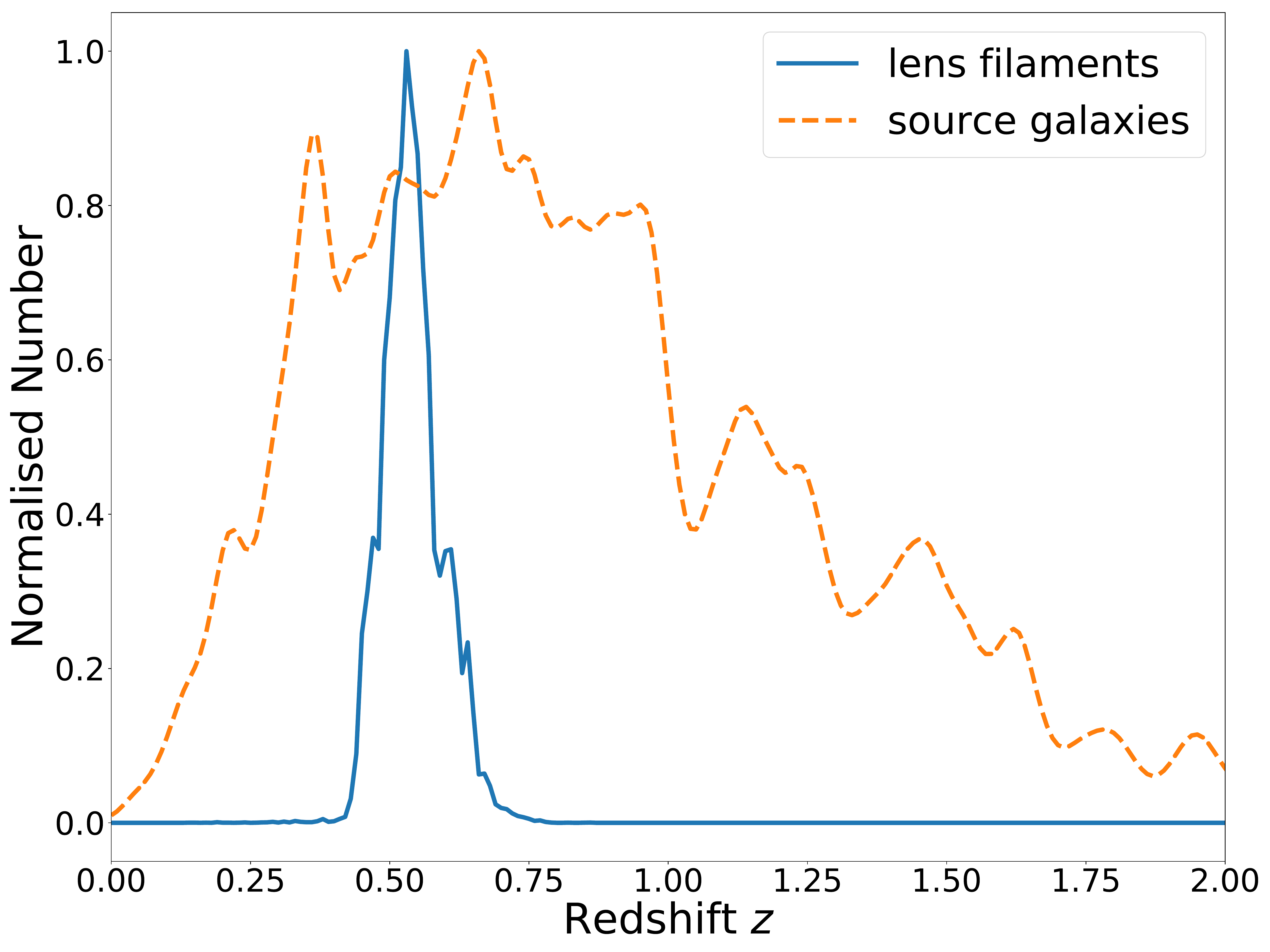}
    \caption{The redshift distribution of lens filaments and source galaxies. Both distributions are normalised for the maximum numbers to be unity. The filament redshift is defined to be a mean redshift of pair galaxies. The source population is much deeper than that of lens filaments, which enables us to perform high signal-to-noise filament lensing measurement.}
    \label{fig:redshift_distribution}
\end{figure}

We basically follow the process described in \citet{Clampitt:2016} to create a galaxy pair sample. We first apply a spatial cut to select CMASS galaxies within the HSC first year full-depth full-color footprints defined in \citet{Mandelbaum:2018a}, which leaves {\rm 14,422} CMASS galaxies. We then apply the line-of-sight distance cut $\left|\Delta \chi\right| < 6~\mpch$ and the physical transverse separation cut $6\mpch < R < 14\mpch$. As a result we obtain 70,210 pairs over the HSC first-year fields. We define the position and redshift of the pair as the avarage of the CMASS galaxy positions and redshifts, respectively. The number of pairs is larger than that of galaxies since a galaxy can be shared among multiple pairs. Figure~\ref{fig:separation_distribution} shows the distribution of transverse separation between the CMASS pairs. Figure~\ref{fig:redshift_distribution} shows the redshift distribution of the CMASS pairs. In our analysis, we use the weight provided by the BOSS LSS catalogue, i.e., $w_l = w_{\rm seeing}w_{\rm star}(w_{\rm noz}+w_{\rm cp}-1)$, where $w_{\rm seeing}$ accounts for the variation in number density due to local seeing, $w_{\rm star}$ accounts for the contamination from stars which is a function of stellar density, $w_{\rm noz}$ accounts for the redshift failure, and $w_{\rm cp}$ accounts for the fibre collision. Note that we do not use the FKP weight $w_{\rm FKP}$, since we do not need to optimise galaxy clustering measurement when measuring the filament lensing signal. We define the pair weight as a product of the weights of the CMASS galaxies 

We also create random pairs using the random catalogue associated with the CMASS LSS catalogue, following the same redshift and separation cuts as the galaxy pairs. We divide the random catalogue such that each sample has the same number of points as the CMASS galaxies, which results in 96 samples of random pairs. The random pairs are used for correcting observational systematics as described in Section~\ref{sec:measurement_filament_lensing}.

For null tests, we create a separated galaxy pair catalogue with CMASS galaxies. Since galaxies largely separated along the line-of-sight are not expected to have filament between them, we select galaxies with the line-of-sight distance cut $ 40~\mpch <\left|\Delta \chi\right| < 60~\mpch$, while keeping the same projected physical transverse separation $6\mpch < R < 14\mpch$ as our pair sample described above.

\subsection{HSC Source Galaxies}
\label{sec:data_hsc}
We use the HSC first-year galaxy shape catalogue as sources to measure filament lensing signal. The details of galaxy shape catalogue is described in \citet{Mandelbaum:2018a} and \citet{Mandelbaum:2018b}, and thus we describe a summary of these papers.

The HSC first-year lensing catalogue is based on the data taken during March 2014 through April 2016 with about 90 nights in total. The total sky coverage is 136.9~deg$^2$ which consists of six distinct fields named GAMA09H, GAMA15H, HECTOMAP, VVDS, WIDE12H, and XMM, and fully overlaps with the BOSS footprint.

The shapes are measured using co-added $i$-band images with a mean seeing FWHM 0.58$^{\prime\prime}$ and 5$\sigma$ point-source detection limit $i_{\rm lim}\sim26$. Despite of the conservative magnitude cut $i<24.5$ to achieve robust shape calibration, the weighted number density of source galaxies is 21.8 arcmin$^{-2}$. The galaxy shapes $(e_1, e_2)$ are estimated using the re-Gaussianization PSF correction method \citep{Hirata:2003}, which was extensively used and characterised in weak lensing studies with the Sloan Digital Sky Survey (SDSS) data \citep{Mandelbaum:2005, Reyes:2012, Mandelbaum:2013}, and then calibrated with the image simulations \citep{Mandelbaum:2018b}  generated by the open-source galaxy image simulation package  \textsc{GalSim} \citep{Rowe:2015}. The calibration factors consist of the multiplicative factor $m$, which is shared among $e_1$ and $e_2$, and the additive bias for each ellipticity component $(c_1, c_2)$. The image simulations are also used to calibrate other quantities which are necessary for lensing signal measurement, i.e., the per-component rms shapes $e_{\rm rms}$, the measurement noise of galaxy shapes $\sigma_e$, and the inverse variance weight from both $e_{\rm rms}$ and $\sigma_e$. Note that we use the shape catalogue with a star mask called Sirius.

Among the photo-$z$s provided by the HSC first-year catalogue \citep{Tanaka:2018}, we use the photo-$z$ estimated by a machine learning method based on self-organising map called MLZ \citep{Carrasco_Kind:2014}
. This is because MLZ is one of the algorithms which yields the smallest systematic uncertainties in lensing measurement within the redshift range of CMASS galaxies \citep{Miyatake:2019} and is used for creating the mock catalogues which will be described in Section~\ref{sec:data_mock}. The redshift distribution of source galaxies are shown in Fig.~\ref{fig:redshift_distribution}. The large fraction of source galaxies are at higher redshifts compared to filaments, which allows for high signal-to-noise filement lensing measurement.

\subsection{Mock Catalogue}
\label{sec:data_mock}
Predicting a filament lensing signal analytically is not straightforward since filaments are not entirely in the linear regime. Analytical estimation of covariance is not straightforward, either, since our filament lensing estimator, which will be described in Section~\ref{sec:measurement_filament_lensing}, uses a source galaxy multiple times and thus strong correlation between spatial bins are expected. In addition, since the HSC first-year field is only $136.9$~deg$^2$, we expect to have large cosmic variance which can contaminate the lensing signal through the intrinsic scatter of filament properties and fluctuations in large scale structure along the line-of-sight. For these reasons, we use the CMASS and source mock catalogue based on $N$-body dark matter only simulations to estimate the theoretical prediction and covariance of filament lensing signal. When estimating these quantities, we use 108 realisations of mocks, each of which corresponds to the measurement with the real data performed in this paper.

For mock source galaxies, we use the mock catalogue generated by \citet{Shirasaki:2019}. Here we summarise how the source mock catalogue was created. Those who are interested in details should refer to the aforementioned paper. The weak lensing shear in mock catalogue is simulated by full-sky ray-tracing through multiple lens planes \citep{Takahashi:2017} which are constructed from 14 boxes with a different box size. Neighbouring lens planes have the comoving distance interval of $\Delta\chi = 150~\mpch$. Source galaxies are distributed in the light-cone simulations following the spatial distribution extracted from the actual HSC shape catalogue and the photo-$z$ estimated by MLZ. The ellipticity of each galaxy is also extracted from the HSC shape catalogue, but is randomly rotated to vanish coherent shear caused by weak lensing in the actual data. The lensing shear obtained from the ray-tracing simulation is then added to the randomised galaxy shapes to mimic the observed galaxy shapes. Note that the weak lensing shear from ray-tracing simulations includes all light deflections by the density fluctuation along the line-of-sight, and thus can be used to derive covariance including the line-of-sight fluctuations. The mock source catalogue also includes the lensing shear without intrinsic shapes of galaxies, which will be used for estimating a theoretical prediction of our filament lensing signal.

We create our CMASS mocks by populating galaxies in dark matter halos in the N-body simulations used for the ray-tracing simulations, following the halo occupation distribution \citep[HOD;][]{Zheng:2005} estimated from the CMASS galaxy-galaxy clustering and abundance. First, we divide the full CMASS LSS sample into five redshift bins so that the binning scheme should be the same as the lens plane definition in the ray-tracing simulations. When one of the edges of the full CMASS sample, i.e., $z_{\rm min}=0.43$ or $z_{\rm max}=0.70$, falls within a redshift bin, we truncate the bin at the edge. Therefore, our final redshift binning is $z\in [0.43, 0.44]$, $[0.44, 0.51]$, $[0.51, 0.57]$, $[0.57, 0.64]$, and $[0.64, 0.70]$. We then measure the galaxy abundance and the projected galaxy-galaxy clustering for each redshift bin in the same manner as \citet{Miyatake:2015}. While using jackknife covariance for clustering, we assume conservative statistical uncertainties for abundance, i.e., $10\%$ of the measured abundance, to absorb the systematic uncertainties in the CMASS selection function due to the depth variations of the survey. To obtain HOD parameters for the CMASS sample within a given redshift bin, we simultaneously fit the projected galaxy-galaxy clustering and abundance signals with cosmological parameters fixed.
To model the clustering and abundance, we use the model based on \textsc{Dark Emulator}, which is described in Appendix G in \cite{Nishimichi:2018} in detail, without off-centering parameters and incompleteness parameters. We populate central galaxies at the centre of halos and satellite galaxies using the measured HOD. When populating satellites, we assume their radial distribution follows that of dark matter in each host halo. The detail of generation of mock HOD galaxies is found in \citet{Shirasaki:2017}. We then generate mock pairs in the same manner as described in Section~\ref{sec:data_boss}, which will be used to estimate theoretical prediction and covariance in Section~\ref{sec:result_C16} and Section~\ref{sec:measurement_covariance}, respectively.

\section{Measurement}
\label{sec:measurement}
In this paper, we use the filament lensing signal estimator proposed by \citet{Clampitt:2016}. This estimator is computed as follows. First, we stretch the pairs such that the line connecting the pairs has the same length. This stretch is done in the same way to both directions of the connecting line and perpendicular to the connecting line. Second, we align the pairs, set up two-dimensional spatial bins by dividing the stretched coordinates into grid cells, and compute the stacked lensing shear in each bin. Finally, we combine the lensing shear computed in multiple bins to cancel out the contribution from dark matter halos. In the following sections, we describe how we compute the lensing signal in a given two-dimensional bin and explain how the \citet{Clampitt:2016} estimator cancels out the contribution from halos. We also describe how we estimate the covariance of the filament lensing signal.

\subsection{Weak Lensing Measurement}
\label{sec:measurement_weak_lensing}
In this section, we describe how to measure the stacked weak lensing signal at a given 2-dimensional bin $\bf{x}$. Using the HSC galaxy shape catalogue, the stacked excess surface mass density $\langle \Delta\Sigma \rangle$ is estimated as
\begin{equation}
\label{eq:DeltaSigma_bin}
\langle \Delta\Sigma_i\rangle({\bf x})=\frac 1 {1+ \hat{m}} \frac{\sum_{ls\in{\bf x}}\left[ w_{ls} \langle \Sigma^{-1}_{\rm cr} \rangle_{ls}^{-1}\left( e_{i,ls}/2\mathcal{R}-c_{i,ls}\right) \right]}{\sum_{ls\in{\bf x}}w_{ls}},
\end{equation}
where $i$ denotes the shear component, i.e., $i=+$ is parallel to the line connecting two CMASS galaxies and $i=\times$ is the one rotated by 45 degrees against the $+$ component, $ls$ runs over lens-source pairs whose source position falls in the 2-dimensional bin ${\bf x}$, $w_{ls}$ is the weight function of a lens-source pair based on the CMASS weight and source weight written as
\begin{equation}
w_{ls} = w_{l1}w_{l2}\frac{\langle \Sigma_{\rm cr}^{-1}\rangle_{ls}^2}{e_{{\rm rms},ls}^2+\sigma_{e,ls}^2},
\end{equation}
where $w_{l1}$ and $w_{l2}$ is the CMASS weight of each galaxy in a CMASS pair, $\langle \Sigma^{-1}_{\rm cr} \rangle_{ls}$ is the expected critical surface mass density for a lens-source pair, which corrects for dilution of lensing signal due to foreground source contamination, defined as
\begin{equation}
\langle \Sigma^{-1}_{\rm cr} \rangle_{ls} = \int_0^\infty dz \Sigma_{\rm cr}^{-1}(z_l, z) P_s(z),
\end{equation}
where $P_s(z)$ is the normalised probability distribution function of a photo-$z$ of the source galaxy and $\Sigma^{-1}_{\rm cr}(z_l, z_s)$ is defined as
\begin{equation}
\Sigma_{\rm cr}^{-1} \left( z_l,z_s\right) = \frac{4\pi G}{c^2}\frac{D_A\left(z_l\right) D_A\left( z_l,z_s\right)}{D_A\left( z_s\right)},
\end{equation}
for $z_s > z_l$ and $\Sigma_{\rm cr}^{-1} \left( z_l,z_s\right)=0$ for $z_s < z_l$, where $D_A(z)$ is the angular diameter distance at redshift $z$, $\mathcal{R}$ is the shear responsivity which corrects for the non-linear operation in the summation process of galaxy ellipticity \citep[e.g.,][]{Bernstein:2002}
\begin{equation}
\mathcal{R}=1-\frac{\sum_{ls\in {\bf x}} w_{ls}e^2_{{\rm rms},ls}}{\sum_{ls\in {\bf x}} w_{ls}},
\end{equation}
$\hat{m}$ is the weighted mean of the multiplicative bias factor
\begin{equation}
\hat{m} = \frac{\sum_{ls\in {\bf x}} w_{ls}m_{ls}}{\sum_{ls\in {\bf x}} w_{ls}},
\end{equation}
and $c_{i,ls}$ is the additive bias for a lens-source pair $ls$.

In Section~\ref{sec:result_2Dmap}, we will show two-dimensional shear map. To compute stacked reduced shear at bin ${\bf x}$, we use the slightly modified version of Eq~\eqref{eq:DeltaSigma_bin}
\begin{equation}
\langle \gamma_i \rangle({\bf x}) \sim \langle g_i \rangle({\bf x})=\frac 1 {1+ \hat{m}} \frac{\sum_{ls\in{\bf x}}\left[ w^\prime_{ls} \left( e_{i,ls}/2\mathcal{R}-c_{i,ls}\right) \right]}{\sum_{ls\in{\bf x}}w^\prime_{ls}},
\end{equation}
where $w^\prime_{ls}$ denotes the inverse-variance weight optimised for the shacked shear measurement
\begin{equation}
w^\prime_{ls} = \frac{w_{l1}w_{l2}}{e_{{\rm rms},ls}^2+\sigma_{e,ls}^2}.
\end{equation}

\subsection{Filament Lensing Estimator}
\label{sec:measurement_filament_lensing}
\begin{figure}
	\includegraphics[width=\columnwidth]{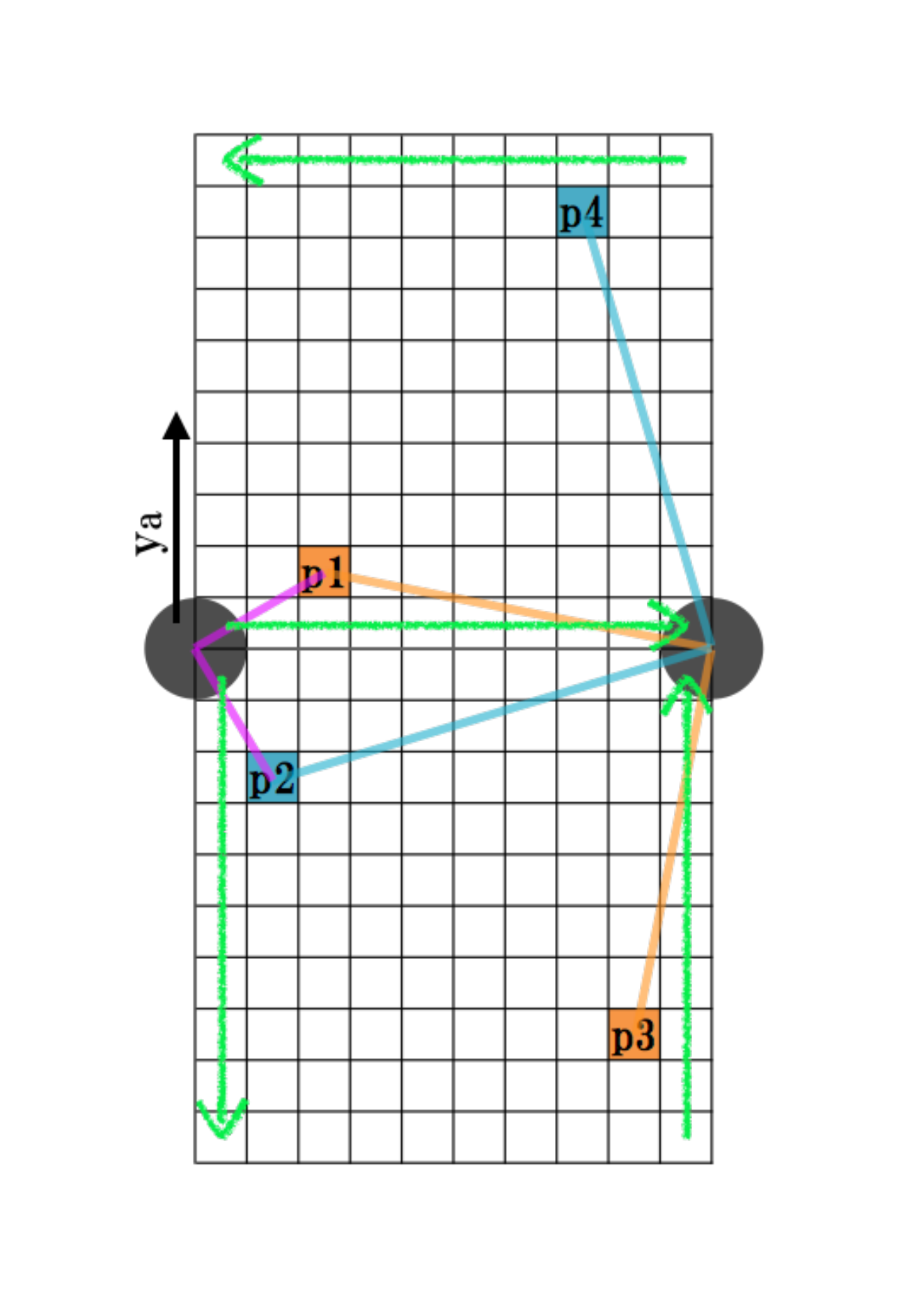}
    \caption{Schematics of the \citet{Clampitt:2016} estimator. Two grey circles denote halos of a pair, and the origin of coordinates is at the centre of the halo on the left-hand side. For the shear at a given position ``p1'', adding the shear at ``p2'' will cancel out the contribution from the halo on the left-hand side. Likewise, adding the shear at ``p3'' and ``p4'' to the shear at ``p1'' and ``p2'' cancels out the contribution from the halo on the right-hand side, respectively. Furthermore, adding the shear at ``p3'' to ``p4'' cancels out the contribution from the halo on the left-hand side. Thus the combination of these four points cancels out all the contribution form the halos. Assuming the translational symmetry of the filament along the line connecting the halos, shears at the positions along the $x$-axis are summed up, which is shown as the green arrow connecting the halos. Taking the combination of four points, shear at the positions along all green arrows are summed up. In addition, assuming the line symmetry of the filament along the connecting line, the positions mirrored against the connecting line are summed up, although it is not shown in this figure for simplicity. The combination of shear at these positions composes the first point of the the \citet{Clampitt:2016} estimator. The other points are computed in the same manner as a function of $y$ as shown in this figure.}
    \label{fig:estimator}
\end{figure}

To extract filament lensing signal from weak lensing shear caused by paired galaxies, we use the estimator proposed by \citet{Clampitt:2016}. This estimator is constructed to cancel out the contribution from lensing signal caused by dark matter halos. Figure~\ref{fig:estimator} shows how this estimator works. Let us suppose to pick up a bin ``p1''. Since the weak lensing shear caused by a spherically symmetric dark matter halo exhibits tangential alignment to the line from the halo centre to the source galaxy\footnote{As described in Appendix of \cite{Clampitt:2016}, elliptical dark matter halo would not significantly affect their filament estimator.}, adding the lensing signal at the point rotated by 90 degrees from ``p1'', which is denoted as ``p2'', cancels out the contribution from the halo on the left-hand side. In addition, both ``p1'' and ``p2'' are also contaminated by the halo on the right-hand side, which can be cancelled out by adding ``p3'' and ``p4'', respectively. Note that the combination of ``p3'' and ``p4'' also cancels out the contribution from the halo on the left-hand side. Assuming the filament has translational and line symmetry along the line connecting the halos, the filament lensing estimator can be written as a function of the distance from the connecting line, which is denoted as $y$,
\begin{eqnarray}
\Delta\Sigma_i^{\rm fil}\left( y_a\right) &\equiv& \frac{1}{80}\sum_{b=1}^{10}\left[ \Delta\Sigma_i\left( x_b,y_a\right) +\Delta\Sigma_i\left( y_a,-x_b\right)\right. \nonumber \\
&&+\Delta\Sigma_i\left( 1-x_b,1-y_a \right)+\Delta\Sigma_i\left( 1-y_a,x_b-1\right) \nonumber \\
&&+\Delta\Sigma_i\left( x_b,-y_a\right)+\Delta\Sigma_i\left( y_a,x_b\right) \nonumber \\
&&\left.+\Delta\Sigma_i\left( 1-x_b,y_a-1\right)+\Delta\Sigma_i\left( 1-y_a,1-x_b\right)\right].
\label{eq:DeltaSigma_filament}
\end{eqnarray}
Here, the centre of halo on the left-hand side is taken as the origin of the stretched coordinates and the distance between two halos is normalised to unity. We divide the connecting line into 10 bins. The each term in the summation on the right-hand side corresponds to Eq.~\eqref{eq:DeltaSigma_bin} where we omit $\langle \rangle$ for simplicity. For example, when $a=2$ and $b=3$, the first, second, third, and fourth term corresponds to ``p1'', ``p2'', ``p4'', and ``p3.''. The summation for the first $y$ bin is denoted as four arrows in Figure~\ref{fig:estimator}, which is denoted as the summation of the first four terms in Eq.~\eqref{eq:DeltaSigma_filament}. Under the assumption of line symmetry, the shear at positions mirrored against the connecting line are also in the summation. These shears correspond to the last four terms in Eq.~\eqref{eq:DeltaSigma_filament}. Note that we take the range of $a$ as $1 \leq a \leq 5$, since $6 \leq a \leq 10$ are already included in calculating the estimator (see the top arrow in Fig.~\ref{fig:estimator} for the $a=1$ case).

We also subtract filament signals estimated with the random pairs from Eq.~\eqref{eq:DeltaSigma_filament}. Although the random pairs do not exhibit any filament signal, the random signal subtraction corrects for observational systematics, such as imperfect PSF correction and imperfect shape noise cancellation at the edge of survey fields. 

\subsection{Covariance}
\label{sec:measurement_covariance}
\begin{figure*}
	\includegraphics[width=\columnwidth]{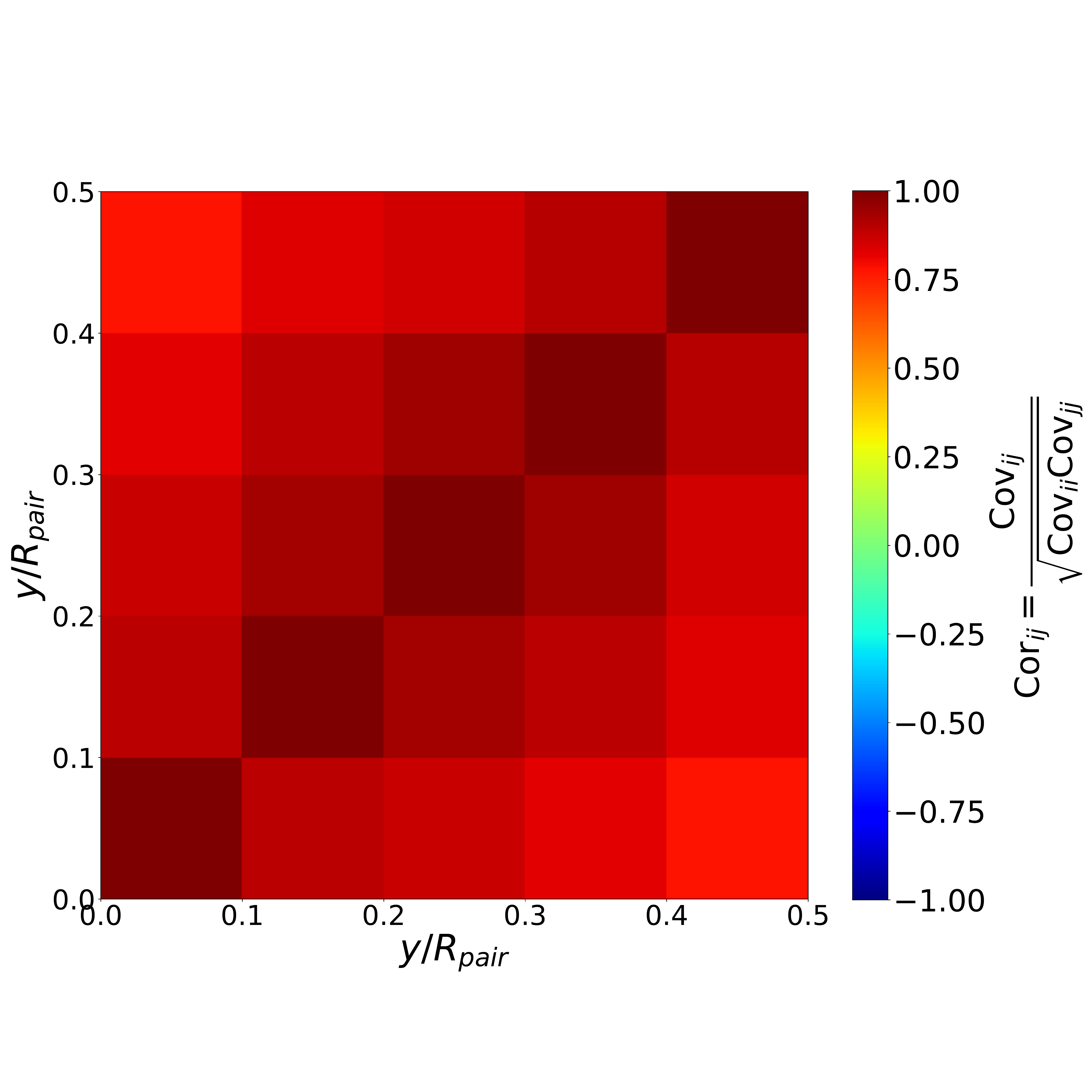}
	\includegraphics[width=\columnwidth]{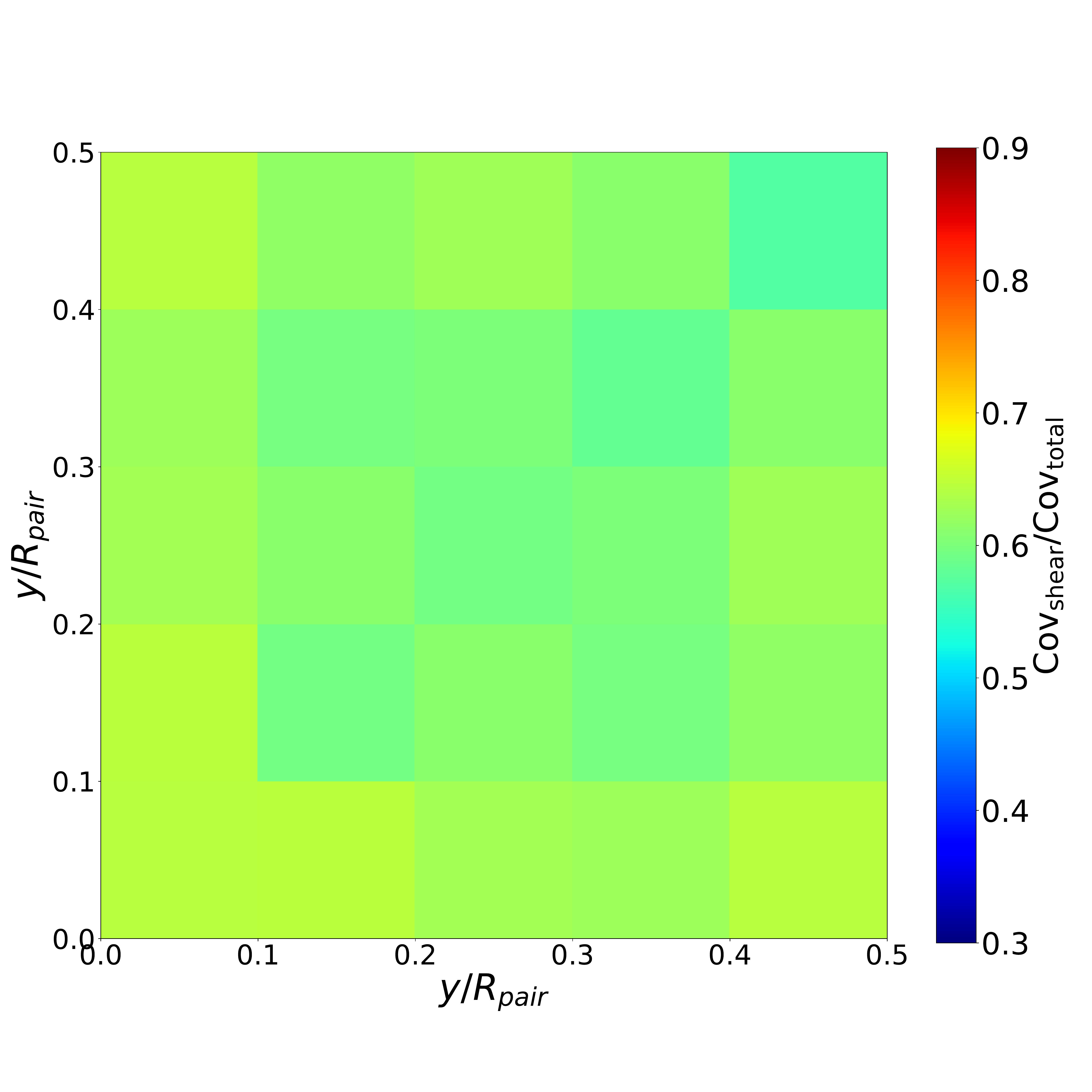}
    \caption{{\it left panel:} Correlation matrix of our filament lensing measurement. The covariance is estimated from from 108 mock realisations. The correlation matrix is higher than 0.77, which means our filament lensing measurement is highly correlated due to the multiple use of source galaxies, the intrinsic scatter of filament properties, and the fluctuations in the projected large scale structure. {\it right panel:} Ratio between covariance estimated from the pure lensing shear in the mock catalogue to the total covariance including intrinsic galaxy shape noise. The covariance from the intrinsic scatter of filament properties and fluctuations in the projected large scale structure occupies 60\% of the covariance, which means our measurement is no longer limited solely by the shape noise.}
    \label{fig:covariance}
\end{figure*}

Estimating the covariance of the filament lensing estimator is not straightforward, since the filament signal at different bins is expected to be significantly correlated due to intrinsic scatter of filament properties, large scale structure projected along the line-of-sight, and multiple use of source galaxies in the stacking process. Using the mock catalogue of CMASS galaxies and source galaxies populated in the ray-tracing simulations described in Section~\ref{sec:data_mock}, we can estimate the covariance naturally taking into account these effects. More precisely, we estimate the filament lensing covariance as
\begin{eqnarray}
{\rm Cov}_i(y_a, y_b)&=&\frac{1}{N_r-1}\nonumber\\
&&\sum_r \left[\Delta\Sigma^{{\rm fil,mock}, r}_i(y_a)-\overline{\Delta\Sigma^{{\rm fil,mock}}_i}(y_a)\right] \nonumber\\
&&\times\left[\Delta\Sigma^{{\rm fil,mock}, r}_i(y_b)-\overline{\Delta\Sigma^{{\rm fil,mock}}_i}(y_b)\right],
\end{eqnarray}
where $i$ denotes the $+$ or $\times$ component, $r$ runs the 108 mock realisations, $\Delta\Sigma^{{\rm fil,mock}, r}_i(y_a)$ is the filament lensing estimator of $r$-th mock realisation, and $\overline{\Delta\Sigma^{{\rm fil,mock}}_i}(y)$ is the estimator averaged over the 108 mock realisations. The correlation matrix of the estimated covariance which is defined as
\begin{equation}
{\rm Cor}_i(y_a, y_b) = \frac{{\rm Cov}_i(y_a, y_b)}{\sqrt{{\rm Cov}_i(y_a, y_a){\rm Cov}_i(y_b, y_b)}}.
\end{equation}
The correlation matrix for the $+$ component is shown in the left panel of Fig.~\ref{fig:covariance}, which exhibits the strong correlation between bins beyond neighbours.

In addition to the observed ellipticity, the mock source galaxy catalogue has the pure weak lensing shear from the ray-tracing simulations. This allows us to estimate covariance without intrinsic shape noise of source galaxies. The right panel of Fig.~\ref{fig:covariance} shows the ratio of the mock shear covariance to the total covariance. The mock shear covariance occupies about 60 percent of the total covariance and is almost constant over both diagonal and off-diagonal components. This means that our filament lensing measurement is no longer limited by the intrinsic shape noise of source galaxies. Rather, the shape noise is comparable to the combination of other sources of covariance which is not relevant to observational conditions, i.e., the fluctuation of large scale structure along the line of sight and the intrinsic scatter of filament properties. This is made possible by the source number density of the HSC survey which is the highest among the ongoing weak lensing surveys.

\section{Results}
\label{sec:result}
\subsection{Filament Lensing Signal}
\label{sec:result_C16}
\begin{figure}
	\includegraphics[width=\columnwidth]{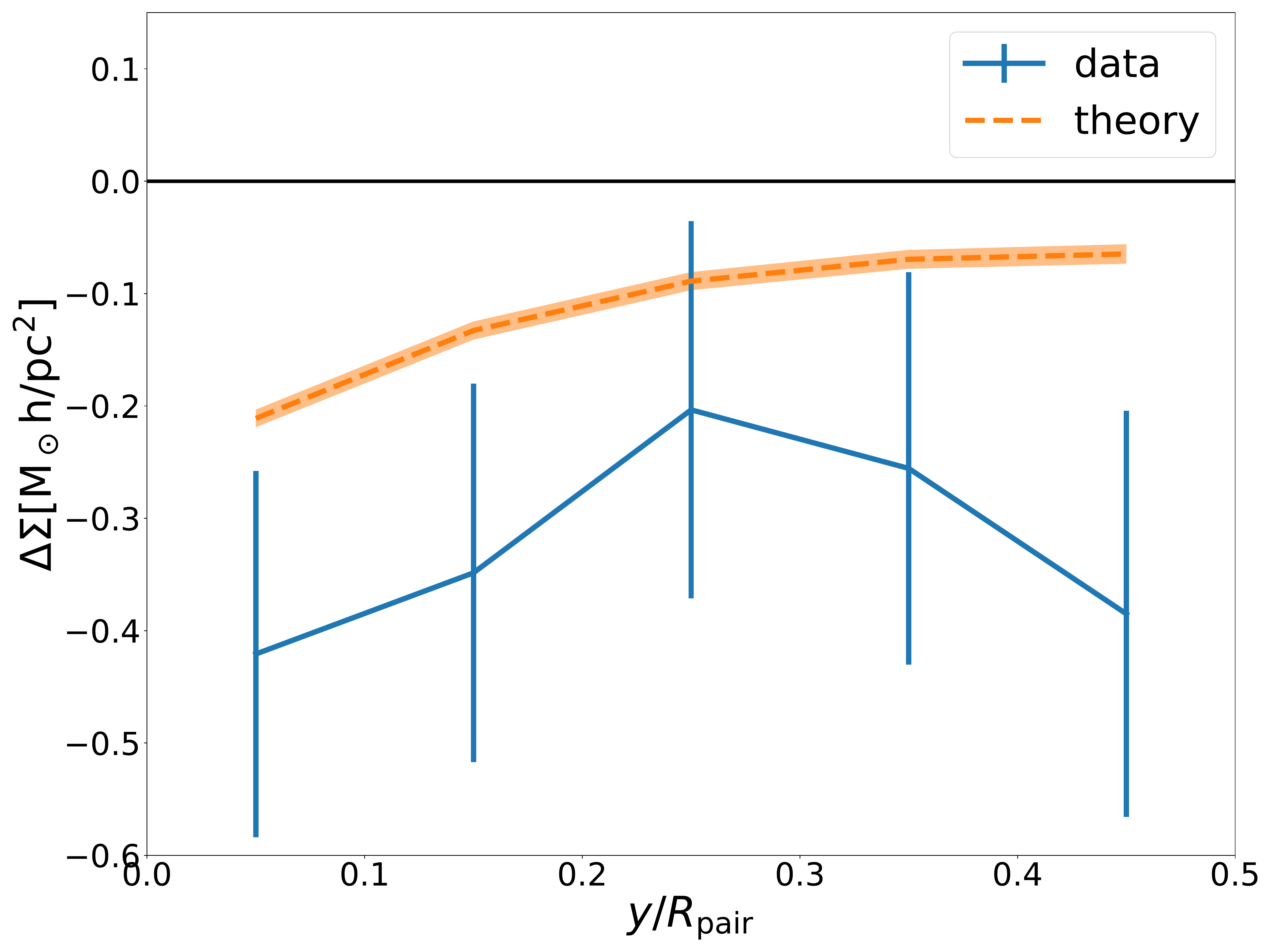}
    \caption{Result of our filament lensing measurement and theoretical prediction. The light-blue solid line denotes our filament lensing measurement of CMASS galaxy pairs. The covariance is estimated from 108 realisations of mock measurements. The filament lensing is detected at 3.9$\sigma$ significance. The orange dashed line is the theoretical prediction obtained by averaging 108 mock measurements. When estimating the theoretical prediction, we use pure weak lensing shear in mocks, i.e., shear without shape noise, to efficiently reduce the statistical uncertainties. The orange band denotes the statistical uncertainties of the averaged mock measurements. The $\chi^2$ value and $p$-value between the measurement and theoretical prediction is 8.1 with the degree of freedom of 5 and 0.15, respectively. Although the data points are consistently smaller than theoretical prediction, the small $\chi^2$ value still suggests the data is consistent with theory. This is because the the filament estimators between the spatial bins are highly correlated.}
    \label{fig:Clampitt}
\end{figure}

Figure~\ref{fig:Clampitt} shows our filament lensing measurement of the CMASS pairs with the error bars estimated by the 108 realisations of mock catalogue. We detect the filament signal at 3.9$\sigma$ significance against null hypothesis. As described in Section~\ref{sec:measurement_covariance}, the intrinsic shape noise of source galaxies is already comparable to the combination of other components in the covariance. Therefore, increasing the number density of galaxies would not help to increase signal-to-noise ratio. Rather, increasing survey area will reduce both the shape noise and other components in the covariance.

The theoretical prediction estimated from the weak lensing shear recorded in the mock catalogues is also shown in Fig.~\ref{fig:Clampitt}. The $\chi^2$ value of the data against the theoretical prediction is $\chi^2=8.1$ with the degree of freedom of 5, which leads to the $p$-value of $0.15$. Hence, our measurement is consistent with the theoretical prediction. Our data points are consistently smaller from the theoretical prediction, and readers may think our $\chi^2$ value is too small. This is due to the fact that the filament lensing signals between bins are highly correlated as shown in Section~\ref{sec:measurement_covariance}.

\subsection{Null Tests}
\label{sec:measurement_null_tests}
To confirm our analysis is not dominated by systematic uncertainties we perform null tests by measureing the $\times$ mode of the CMASS pair lensing signal and the $+$ and $\times$ modes of the separated CMASS pair lensing signal.

\begin{figure}
	\includegraphics[width=\columnwidth]{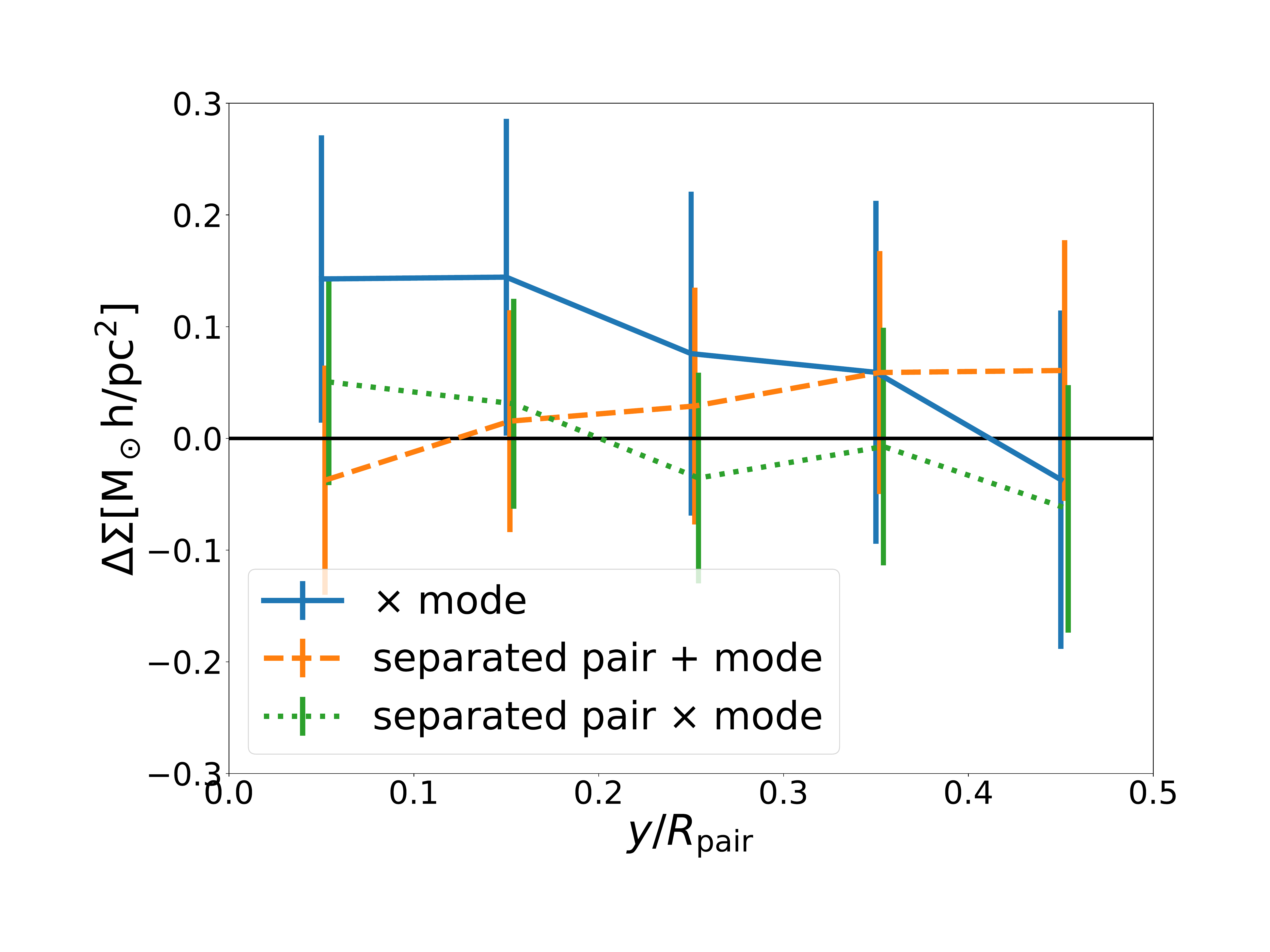}
    \caption{Summary plot of null tests. The light-blue solid line denotes the $\times$-mode lensing signal of the CMASS pairs. The orange dashed line and green dotted line denotes the $+$-mode and $x$-mode lensing signal of CMASS pairs whose separation along the line-of-sight is $40\mpch < \left|\Delta\chi\right|  < 60\mpch$, respectively. The covariance is estimated from the measurements on 108 realisations of mock simulations which are performed exactly in the same manner as on the data. All the null tests are passed as summarised in Table~\ref{table:null_test}. Note that the orange dashed line and green dotted line are slightly shifted along the $x$-axis for illustrative purposes.}
    \label{fig:systematics}
\end{figure}

We expect the $\times$ mode of the filament lensing estimator is consistent with zero due to the translational symmetry of the filament along the line connecting the pair. The $\times$-mode signal is shown in Fig.~\ref{fig:systematics} and the $\chi^2$ value and $p$-value are summarised in Table~\ref{table:null_test}, which confirms that the the $\times$ mode is consistent with null.

We also expect both the $+$ mode and the $\times$ mode of the separated pair signal are consistent with null, since the separated pair is defined such that their line-of-sight distance is $ 40~\mpch<\left|\Delta \chi\right| < 60~\mpch$ as described in Section~\ref{sec:data_boss} and we do not expect massive filaments exist between the halos. These signals are shown in Fig.~\ref{fig:systematics} and the $\chi^2$ values and $p$-values are summarised in Table~\ref{table:null_test}. We also confirm that separated pair signals are consistent with null.

\begin{table}
	\centering
	\caption{Summary table of null tests. Note that we use the signal after the random signal correction.}
	\label{tab:example_table}
	\begin{tabular}{lcc} 
		\hline
		null test & $\chi^2$/dof & $p$-value\\
		\hline
		$\times$ mode & 7.27/5 & 0.20\\
		separated pair $+$ mode & 2.34/5 & 0.80\\
		separated pair $\times$ mode & 6.05/5 & 0.30\\
		\hline
	\end{tabular}
	\label{table:null_test}
\end{table}

\subsection{Two-dimensional Shear and Convergence Map}
\label{sec:result_2Dmap}
\begin{figure*}
	\includegraphics[width=\columnwidth]{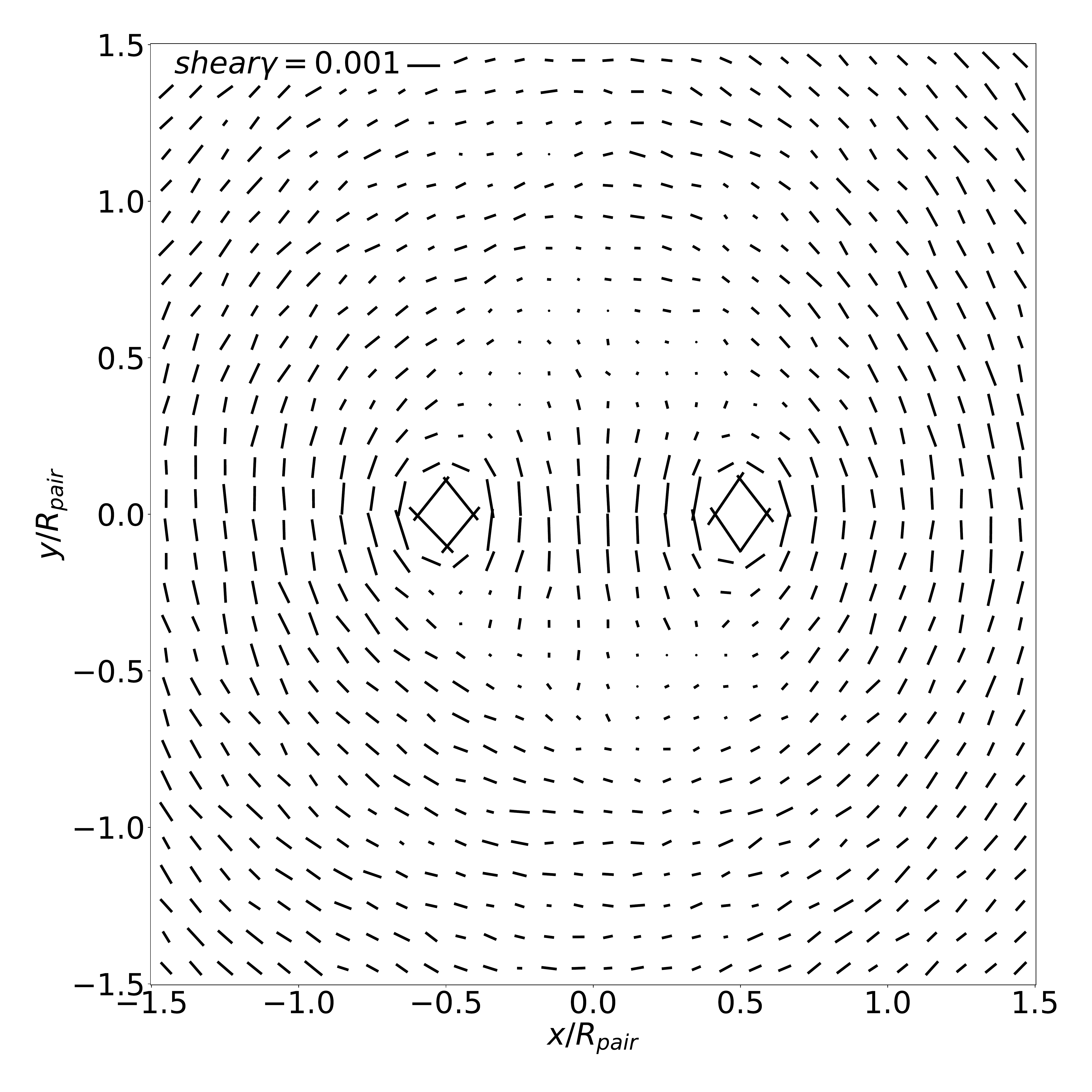}
    \includegraphics[width=\columnwidth]{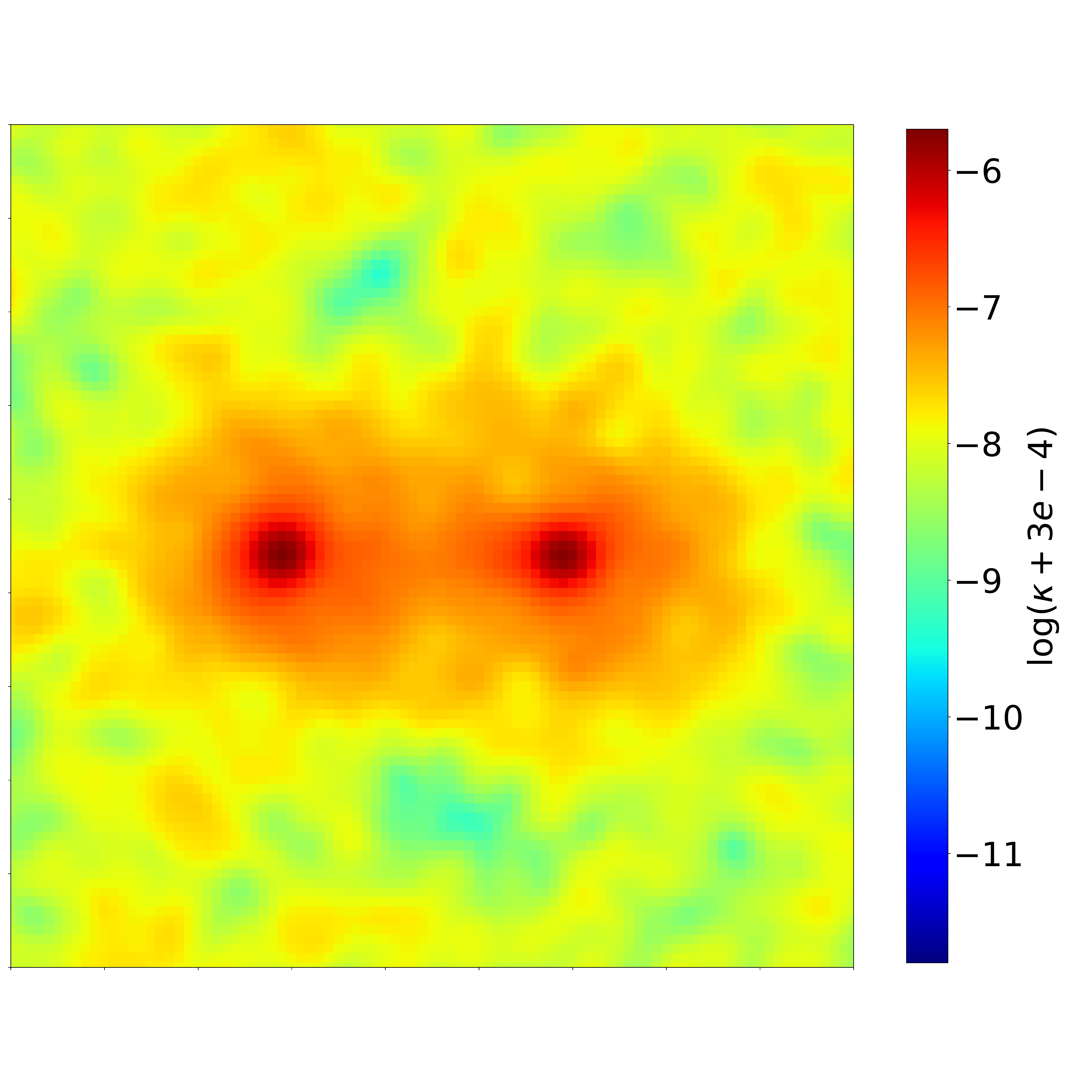}
    \caption{{\it left panel:} Two-dimensional shear map of the stacked CMASS galaxy pairs. The shear between halos is composed of the tangential shear from both halos and the shear from filament which is perpendicular to the line connecting halos. Although the latter is unclear in this map, the filament signal is observed in the \citet{Clampitt:2016} estimator as shown in Fig.~\ref{fig:Clampitt}. The tangential shear surrounding the pair is from the large scale structure associated with the pair, which can be expressed as the galaxy-galaxy-shear correlation function. {\it right panel:} Two-dimensional convergence map of the stacked CMASS galaxy pairs computed from the shear map. The colour scale shows $\log(\kappa+10^{-3})$, where we add the offset to $\kappa$ to avoid a negative argument in the logarithmic function. The convergence map supports the ``thick'' filament model derived from $N$-body simulations.}
    \label{fig:2d_map}
\end{figure*}

The left panel of Fig.~\ref{fig:2d_map} shows two-dimensional stacked shear map of our CMASS pair sample. The vertical shear between CMASS halos are the sum of the tangential shear from CMASS halos and filament shear. Just from this figure it is not clear if there is a filament between halos. This demonstrates that it is hard to distinguish the filaments component from dark matter halos in two-dimensional shear map. However, the \citet{Clampitt:2016} estimator nicely eliminates the contributions from halos, and thus isolates the signal from the filaments, as described in Section~\ref{sec:result_C16}. In Fig.~\ref{fig:2d_map}, we observe lensing shear surrounding the pair at large scales. This is part of the galaxy-galaxy-shear correlation function, which was measured by \cite{Simon:2013}.

From this shear map, we compute the convergence map, which is shown in the right panel of Fig.~\ref{fig:2d_map}. In doing so, we use the prescription by \cite{Kaiser:1993} with the convolution kernel adopted in \cite{Miyazaki:2015}, i.e.,
\begin{equation}
\kappa({\bf x}) = \int d^2 y \gamma_t({\bf y};{\bf x}))Q_G(\left|{\bf y}\right|),
\end{equation}
where $\gamma_t$ is the tangential component of the shear at the position ${\bf y}$ with respect to ${\bf x}$ and $Q(\left|{\bf x}\right|)$ is the convolution kernel
\begin{equation}
Q_G(x)=\frac{1}{\pi x^2}\left[1-\left(1+\frac{x^2}{x_G^2}\right)\exp\left(-\frac{x^2}{x_G^2}\right) \right].
\end{equation}
We choose the width of kernel as $x_G=R_{\rm pair}/12$. In this convergence map, we can clearly see the diffuse filament between circularly symmetric galaxy halos. This independently supports the ``thick'' filament predicted by the $N$-body simulations rather than the ``thin'' filament modelled by a line of small halos.

\subsection{Comparison with Previous Filament Lensing Measurements}
\label{sec:result_comparion}
The first detection of the filament between massive galaxy pairs was made by \citet{Clampitt:2016} using SDSS LRG pairs at a typical redshift $z\sim0.25$ and SDSS source galaxies, where they used the exactly same estimator as ours. We significantly extend the redshift range of galaxy pairs to $z\sim0.55$. The significance level of their detection in their measurement was 4.5$\sigma$, which is slightly higher than ours despite of their low number density of source galaxies ($\sim0.5$~arcmin$^{-2}$). This is because their survey area ($\sim8000$~deg$^2$) is much larger than ours ($\sim140$~deg$^2$), and thus their measurement is not limited by the large scale structure noise. This in turn means that we will soon be able to measure the lensing signal at a higher signal-to-noise ratio as the HSC survey area increases. The amplitude of their filament signal is as twice as large as ours. This is likely due to the difference in halo mass between the LRG and CMASS sample, i.e., the CMASS sample is less massive than LRG, which results in less massive filaments. This might be also because the structure growth is larger at $z=0.25$ than those at $z=0.55$, so the filaments may have more masses even if the mass of the halos are the same. We may check this effect by using mock simulations, but this is beyond the scope of this paper.

\citet{Epps:2017} reported the filament measurement of galaxy pairs of the combined BOSS LOWZ and CMASS sample with the mean redshift of $\langle z\rangle \sim0.42$, using the CFHTLenS source galaxies \citep{Heymans:2012}. Using the two-dimensional convergence map where the contribution from halos was removed by using their ``control'' sample, they claimed the detection at 5$\sigma$ significance. They explored modelling the filament signal by the galaxy-galaxy-convergence three point correlation function, and found reasonable agreement with their measurement. In contrast, our pair sample purely consists of the CMASS sample which pushed the mean redshift to $\langle z\rangle \sim0.55$. We also derive the more robust theoretical prediction using the mock CMASS pairs and source galaxies based on ray-tracing simulations.

\cite{He:2018} reported the first detection of filament lensing signal imprinted in CMB by cross-correlating the filament map derived from the CMASS galaxies \citep{Chen:2016} with the Planck CMB convergence map \citep{Planck2013XVII} in Fourier space. Note that their filament sample is intrinsically different from our sample. They observed filaments where CMASS galaxies reside whereas we observe filaments between CMASS pairs, and thus their filament sample is expected to be more massive than ours.

The main focus of \citet{deGraaff:2019} was probing missing baryons through thermal Sunyaev-Zel'dovich effect \citep[tSZ;][]{Sunyaev:1969, Sunyaev:1972} from filaments between CMASS pairs, but they also measured lensing signal using the Planck convergence map to investigate the connection between gas and dark matter filaments. Their two-dimensional convergence measurement yielded 1.9$\sigma$, which is smaller than our measurement despite of their use of the entire CMASS sample without any spatial cut. This fact demonstrates the statistical power of the HSC survey enabled by the high number density or source galaxies. In addition, HSC source galaxies has the lensing kernel which matches to the CMASS sample better than CMB lensing.

\section{Conclusions}
\label{sec:conclusions}
We have measured weak gravitational lensing caused by filamentary structures between the CMASS galaxy pairs using the Subaru HSC first-year galaxy shape catalogue. Desite of the small overlap between BOSS and HSC ($\sim 140$~deg$^2$), the high number density of the HSC shape catalogue ($\sim 20$~arcmin$^{-2}$) enabled us to measure the filament lensing signal at 3.9$\sigma$ significance, which is the highest signal-to-noise measurement of the filament between high-redshift ($z\sim0.55$) galaxy-scale halos to date. We have used the \cite{Clampitt:2016} estimator to cancel out the contribution from the dark matter halos and extract the signal from the filaments. We have found that the covariance is highly correlated and the contribution from the combination of the intrinsic scatter of filament properties and the fluctuations in large scale structure is already comparable to the contribution from the intrinsic shape noise from source galaxies. This demonstrates the statistical power of the HSC survey. 
We expect the signal-to-noise ratio will rapidly increase as the survey field grows. At the end of the HSC survey, the signal-to-noise ratio of our measurement will be improved by the factor of three, which will allows us to do more detailed studies such as the dark matter distribution and redshift evolution of filaments. We have observed negative value of the \cite{Clampitt:2016} estimator, which supports the ``thick'' filament model and consistent with what \citet{Clampitt:2016} found in the LRG pair sample. We have confirmed this fact by drawing the two-dimensional convergence map. 

\citet{deGraaff:2019} performed a pioneering work to combine multi-wavelength data to study the connection between light, gas, and dark matter within filaments. In coming years, as various surveys for imaging, spectroscopic, X-ray, and SZ will collect more data, more detailed studies to investigate such astrophysical properties of filaments will become possible.

Using the ongoing and upcoming spectroscopic surveys such as SDSS-IV/eBOSS \citep{Dawson:2016}, Dark Energy Spectroscopic Instrument \citep[DESI;][]{DESI:2016}, and Subaru Prime Focus Pectrograph \citep[PFS;][]{Takada:2014}, we can extend the redshift range of a filament sample. In addition, the upcoming imaging surveys such as Large Synoptic Survey Telescope \citep[LSST;][]{LSST:2009}, Euclid \citep{Laureijs:2011}, Wide-field Infrared Survey Telescope \citep[WFIRST;][]{Spergel:2015} will 
enable us to pursue the weak lensing measurement of these filaments with even higher signal-to-noise ratio.

\section*{Acknowledgements}
The Hyper Suprime-Cam (HSC) collaboration includes the astronomical communities of Japan and Taiwan, and Princeton University.  The HSC instrumentation and software were developed by the National Astronomical Observatory of Japan (NAOJ), the Kavli Institute for the Physics and Mathematics of the Universe (Kavli IPMU), the University of Tokyo, the High Energy Accelerator Research Organization (KEK), the Academia Sinica Institute for Astronomy and Astrophysics in Taiwan (ASIAA), and Princeton University.  Funding was contributed by the FIRST program from Japanese Cabinet Office, the Ministry of Education, Culture, Sports, Science and Technology (MEXT), the Japan Society for the Promotion of Science (JSPS),  Japan Science and Technology Agency  (JST),  the Toray Science  Foundation, NAOJ, Kavli IPMU, KEK, ASIAA,  and Princeton University.

The Pan-STARRS1 Surveys (PS1) have been made possible through contributions of the Institute for Astronomy, the University of Hawaii, the Pan-STARRS Project Office, the Max-Planck Society and its participating institutes, the Max Planck Institute for Astronomy, Heidelberg and the Max Planck Institute for Extraterrestrial Physics, Garching, The Johns Hopkins University, Durham University, the University of Edinburgh, Queen's University Belfast, the Harvard-Smithsonian Center for Astrophysics, the Las Cumbres Observatory Global Telescope Network Incorporated, the National Central University of Taiwan, the Space Telescope Science Institute, the National Aeronautics and Space Administration under Grant No. NNX08AR22G issued through the Planetary Science Division of the NASA Science Mission Directorate, the National Science Foundation under Grant No. AST-1238877, the University of Maryland, and Eotvos Lorand University (ELTE).
 
This paper makes use of software developed for the Large Synoptic Survey Telescope. We thank the LSST Project for making their code available as free software at http://dm.lsst.org.

Based in part on data collected at the Subaru Telescope and retrieved from the HSC data archive system, which is operated by the Subaru Telescope and Astronomy Data Center at National Astronomical Observatory of Japan.

HM acknowledges the support from JSPS KAKENHI Grant Number JP18H04350, JP18K13561, and JP19H05100.
MS acknowledges the support in part from by JSPS KAKENHI Grant Number JP18H04358. Numerical computations were in part carried out on Cray XC50 at Center for Computational Astrophysics, National Astronomical Observatory of Japan.
AJN acknowledges the support in part by MEXT KAKENHI Grant Number 15H05890.




\bibliographystyle{mnras}
\bibliography{main} 







\bsp	
\label{lastpage}

\end{document}

%% file: commands.tex
\newcommand{\mpch}{\>h^{-1}{\rm {Mpc}}}

\newcommand{\kms}{\>{\rm km}\,{\rm s}^{-1}}

\def\gcm3{\mathrm{g} / \mathrm{cm}^3}

\def\gtsima{$\; \buildrel > \over \sim \;$}
\def\ltsima{$\; \buildrel < \over \sim \;$}
\def\prosima{$\; \buildrel \propto \over \sim \;$}
\def\gsim{\lower.7ex\hbox{\gtsima}}
\def\lsim{\lower.7ex\hbox{\ltsima}}
\def\simgt{\lower.7ex\hbox{\gtsima}}
\def\simlt{\lower.7ex\hbox{\ltsima}}
\def\simpr{\lower.7ex\hbox{\prosima}}

\def\bm{{\bf m}}

%% file: main.bbl
\begin{thebibliography}{}
\makeatletter
\relax
\def\mn@urlcharsother{\let\do\@makeother \do\$\do\&\do\#\do\^\do\_\do\%\do\~}
\def\mn@doi{\begingroup\mn@urlcharsother \@ifnextchar [ {\mn@doi@}
  {\mn@doi@[]}}
\def\mn@doi@[#1]#2{\def\@tempa{#1}\ifx\@tempa\@empty \href
  {http://dx.doi.org/#2} {doi:#2}\else \href {http://dx.doi.org/#2} {#1}\fi
  \endgroup}
\def\mn@eprint#1#2{\mn@eprint@#1:#2::\@nil}
\def\mn@eprint@arXiv#1{\href {http://arxiv.org/abs/#1} {{\tt arXiv:#1}}}
\def\mn@eprint@dblp#1{\href {http://dblp.uni-trier.de/rec/bibtex/#1.xml}
  {dblp:#1}}
\def\mn@eprint@#1:#2:#3:#4\@nil{\def\@tempa {#1}\def\@tempb {#2}\def\@tempc
  {#3}\ifx \@tempc \@empty \let \@tempc \@tempb \let \@tempb \@tempa \fi \ifx
  \@tempb \@empty \def\@tempb {arXiv}\fi \@ifundefined
  {mn@eprint@\@tempb}{\@tempb:\@tempc}{\expandafter \expandafter \csname
  mn@eprint@\@tempb\endcsname \expandafter{\@tempc}}}

\bibitem[\protect\citeauthoryear{{Aihara} et~al.,}{{Aihara}
  et~al.}{2018a}]{Aihara:2018a}
{Aihara} H.,  et~al., 2018a, \mn@doi [\pasj] {10.1093/pasj/psx066}, \href
  {http://adsabs.harvard.edu/abs/2018PASJ...70S...4A} {70, S4}

\bibitem[\protect\citeauthoryear{{Aihara} et~al.,}{{Aihara}
  et~al.}{2018b}]{Aihara:2018b}
{Aihara} H.,  et~al., 2018b, \mn@doi [\pasj] {10.1093/pasj/psx081}, \href
  {http://adsabs.harvard.edu/abs/2018PASJ...70S...8A} {70, S8}

\bibitem[\protect\citeauthoryear{{Alam} et~al.,}{{Alam}
  et~al.}{2015}]{Alam:2015}
{Alam} S.,  et~al., 2015, \mn@doi [\apjs] {10.1088/0067-0049/219/1/12}, \href
  {http://adsabs.harvard.edu/abs/2015ApJS..219...12A} {219, 12}

\bibitem[\protect\citeauthoryear{{Bernstein} \& {Jarvis}}{{Bernstein} \&
  {Jarvis}}{2002}]{Bernstein:2002}
{Bernstein} G.~M.,  {Jarvis} M.,  2002, \mn@doi [\aj] {10.1086/338085}, \href
  {http://adsabs.harvard.edu/abs/2002AJ....123..583B} {123, 583}

\bibitem[\protect\citeauthoryear{{Carrasco Kind} \& {Brunner}}{{Carrasco Kind}
  \& {Brunner}}{2014}]{Carrasco_Kind:2014}
{Carrasco Kind} M.,  {Brunner} R.~J.,  2014, \mn@doi [\mnras]
  {10.1093/mnras/stt2456}, \href
  {http://adsabs.harvard.edu/abs/2014MNRAS.438.3409C} {438, 3409}

\bibitem[\protect\citeauthoryear{{Chen}, {Ho}, {Brinkmann}, {Freeman},
  {Genovese}, {Schneider}  \& {Wasserman}}{{Chen} et~al.}{2016}]{Chen:2016}
{Chen} Y.-C.,  {Ho} S.,  {Brinkmann} J.,  {Freeman} P.~E.,  {Genovese} C.~R.,
  {Schneider} D.~P.,   {Wasserman} L.,  2016, \mn@doi [\mnras]
  {10.1093/mnras/stw1554}, \href
  {https://ui.adsabs.harvard.edu/abs/2016MNRAS.461.3896C} {461, 3896}

\bibitem[\protect\citeauthoryear{{Clampitt}, {Miyatake}, {Jain}  \&
  {Takada}}{{Clampitt} et~al.}{2016}]{Clampitt:2016}
{Clampitt} J.,  {Miyatake} H.,  {Jain} B.,   {Takada} M.,  2016, \mn@doi
  [\mnras] {10.1093/mnras/stw142}, \href
  {http://adsabs.harvard.edu/abs/2016MNRAS.457.2391C} {457, 2391}

\bibitem[\protect\citeauthoryear{{Colless} et~al.,}{{Colless}
  et~al.}{2001}]{Colless:2001}
{Colless} M.,  et~al., 2001, \mn@doi [\mnras]
  {10.1046/j.1365-8711.2001.04902.x}, \href
  {http://adsabs.harvard.edu/abs/2001MNRAS.328.1039C} {328, 1039}

\bibitem[\protect\citeauthoryear{{DESI Collaboration} et~al.,}{{DESI
  Collaboration} et~al.}{2016}]{DESI:2016}
{DESI Collaboration} et~al., 2016, arXiv e-prints, \href
  {https://ui.adsabs.harvard.edu/abs/2016arXiv161100036D} {p. arXiv:1611.00036}

\bibitem[\protect\citeauthoryear{{Dark Energy Survey Collaboration}
  et~al.,}{{Dark Energy Survey Collaboration} et~al.}{2016}]{DES:2016}
{Dark Energy Survey Collaboration} et~al., 2016, \mn@doi [\mnras]
  {10.1093/mnras/stw641}, \href
  {https://ui.adsabs.harvard.edu/abs/2016MNRAS.460.1270D} {460, 1270}

\bibitem[\protect\citeauthoryear{{Dawson} et~al.,}{{Dawson}
  et~al.}{2013}]{Dawson:2013}
{Dawson} K.~S.,  et~al., 2013, \mn@doi [\aj] {10.1088/0004-6256/145/1/10},
  \href {http://adsabs.harvard.edu/abs/2013AJ....145...10D} {145, 10}

\bibitem[\protect\citeauthoryear{{Dawson} et~al.,}{{Dawson}
  et~al.}{2016}]{Dawson:2016}
{Dawson} K.~S.,  et~al., 2016, \mn@doi [\aj] {10.3847/0004-6256/151/2/44},
  \href {http://adsabs.harvard.edu/abs/2016AJ....151...44D} {151, 44}

\bibitem[\protect\citeauthoryear{{Dietrich}, {Werner}, {Clowe}, {Finoguenov},
  {Kitching}, {Miller}  \& {Simionescu}}{{Dietrich}
  et~al.}{2012}]{Dietrich:2012}
{Dietrich} J.~P.,  {Werner} N.,  {Clowe} D.,  {Finoguenov} A.,  {Kitching} T.,
  {Miller} L.,   {Simionescu} A.,  2012, \mn@doi [\nat] {10.1038/nature11224},
  \href {http://adsabs.harvard.edu/abs/2012Natur.487..202D} {487, 202}

\bibitem[\protect\citeauthoryear{{Epps} \& {Hudson}}{{Epps} \&
  {Hudson}}{2017}]{Epps:2017}
{Epps} S.~D.,  {Hudson} M.~J.,  2017, \mn@doi [\mnras] {10.1093/mnras/stx517},
  \href {https://ui.adsabs.harvard.edu/abs/2017MNRAS.468.2605E} {468, 2605}

\bibitem[\protect\citeauthoryear{{Furusawa} et~al.,}{{Furusawa}
  et~al.}{2018}]{Furusawa:2018}
{Furusawa} H.,  et~al., 2018, \mn@doi [\pasj] {10.1093/pasj/psx079}, \href
  {https://ui.adsabs.harvard.edu/abs/2018PASJ...70S...3F} {70, S3}

\bibitem[\protect\citeauthoryear{{He}, {Alam}, {Ferraro}, {Chen}  \& {Ho}}{{He}
  et~al.}{2018}]{He:2018}
{He} S.,  {Alam} S.,  {Ferraro} S.,  {Chen} Y.-C.,   {Ho} S.,  2018, \mn@doi
  [Nature Astronomy] {10.1038/s41550-018-0426-z}, \href
  {http://adsabs.harvard.edu/abs/2018NatAs...2..401H} {2, 401}

\bibitem[\protect\citeauthoryear{{Heymans} et~al.,}{{Heymans}
  et~al.}{2012}]{Heymans:2012}
{Heymans} C.,  et~al., 2012, \mn@doi [\mnras]
  {10.1111/j.1365-2966.2012.21952.x}, \href
  {http://adsabs.harvard.edu/abs/2012MNRAS.427..146H} {427, 146}

\bibitem[\protect\citeauthoryear{{Higuchi}, {Oguri}, {Tanaka}  \&
  {Sakurai}}{{Higuchi} et~al.}{2015}]{Higuchi:2015}
{Higuchi} Y.,  {Oguri} M.,  {Tanaka} M.,   {Sakurai} J.,  2015, arXiv e-prints,
  \href {http://adsabs.harvard.edu/abs/2015arXiv150306373H} {p.
  arXiv:1503.06373}

\bibitem[\protect\citeauthoryear{{Hirata} \& {Seljak}}{{Hirata} \&
  {Seljak}}{2003}]{Hirata:2003}
{Hirata} C.,  {Seljak} U.,  2003, \mn@doi [\mnras]
  {10.1046/j.1365-8711.2003.06683.x}, \href
  {http://adsabs.harvard.edu/abs/2003MNRAS.343..459H} {343, 459}

\bibitem[\protect\citeauthoryear{{Jauzac} et~al.,}{{Jauzac}
  et~al.}{2012}]{Jauzac:2012}
{Jauzac} M.,  et~al., 2012, \mn@doi [\mnras]
  {10.1111/j.1365-2966.2012.21966.x}, \href
  {https://ui.adsabs.harvard.edu/abs/2012MNRAS.426.3369J} {426, 3369}

\bibitem[\protect\citeauthoryear{{Kaiser} \& {Squires}}{{Kaiser} \&
  {Squires}}{1993}]{Kaiser:1993}
{Kaiser} N.,  {Squires} G.,  1993, \mn@doi [\apj] {10.1086/172297}, \href
  {http://adsabs.harvard.edu/abs/1993ApJ...404..441K} {404, 441}

\bibitem[\protect\citeauthoryear{{Kawanomoto} et~al.,}{{Kawanomoto}
  et~al.}{2018}]{Kawanomoto:2018}
{Kawanomoto} S.,  et~al., 2018, \mn@doi [\pasj] {10.1093/pasj/psy056}, \href
  {https://ui.adsabs.harvard.edu/abs/2018PASJ...70...66K} {70, 66}

\bibitem[\protect\citeauthoryear{{Komiyama} et~al.,}{{Komiyama}
  et~al.}{2018}]{Komiyama:2018}
{Komiyama} Y.,  et~al., 2018, \mn@doi [\pasj] {10.1093/pasj/psx069}, \href
  {https://ui.adsabs.harvard.edu/abs/2018PASJ...70S...2K} {70, S2}

\bibitem[\protect\citeauthoryear{{LSST Science Collaboration} et~al.,}{{LSST
  Science Collaboration} et~al.}{2009}]{LSST:2009}
{LSST Science Collaboration} et~al., 2009, arXiv e-prints, \href
  {https://ui.adsabs.harvard.edu/abs/2009arXiv0912.0201L} {p. arXiv:0912.0201}

\bibitem[\protect\citeauthoryear{{Laureijs} et~al.,}{{Laureijs}
  et~al.}{2011}]{Laureijs:2011}
{Laureijs} R.,  et~al., 2011, arXiv e-prints, \href
  {https://ui.adsabs.harvard.edu/abs/2011arXiv1110.3193L} {p. arXiv:1110.3193}

\bibitem[\protect\citeauthoryear{{Mandelbaum} et~al.,}{{Mandelbaum}
  et~al.}{2005}]{Mandelbaum:2005}
{Mandelbaum} R.,  et~al., 2005, \mn@doi [\mnras]
  {10.1111/j.1365-2966.2005.09282.x}, \href
  {http://adsabs.harvard.edu/abs/2005MNRAS.361.1287M} {361, 1287}

\bibitem[\protect\citeauthoryear{{Mandelbaum}, {Slosar}, {Baldauf}, {Seljak},
  {Hirata}, {Nakajima}, {Reyes}  \& {Smith}}{{Mandelbaum}
  et~al.}{2013}]{Mandelbaum:2013}
{Mandelbaum} R.,  {Slosar} A.,  {Baldauf} T.,  {Seljak} U.,  {Hirata} C.~M.,
  {Nakajima} R.,  {Reyes} R.,   {Smith} R.~E.,  2013, \mn@doi [\mnras]
  {10.1093/mnras/stt572}, \href
  {http://adsabs.harvard.edu/abs/2013MNRAS.432.1544M} {432, 1544}

\bibitem[\protect\citeauthoryear{{Mandelbaum} et~al.,}{{Mandelbaum}
  et~al.}{2018a}]{Mandelbaum:2018a}
{Mandelbaum} R.,  et~al., 2018a, \mn@doi [\pasj] {10.1093/pasj/psx130}, \href
  {http://adsabs.harvard.edu/abs/2018PASJ...70S..25M} {70, S25}

\bibitem[\protect\citeauthoryear{{Mandelbaum} et~al.,}{{Mandelbaum}
  et~al.}{2018b}]{Mandelbaum:2018b}
{Mandelbaum} R.,  et~al., 2018b, \mn@doi [\mnras] {10.1093/mnras/sty2420},
  \href {http://adsabs.harvard.edu/abs/2018MNRAS.481.3170M} {481, 3170}

\bibitem[\protect\citeauthoryear{{Miyatake} et~al.,}{{Miyatake}
  et~al.}{2015}]{Miyatake:2015}
{Miyatake} H.,  et~al., 2015, \mn@doi [\apj] {10.1088/0004-637X/806/1/1}, \href
  {http://adsabs.harvard.edu/abs/2015ApJ...806....1M} {806, 1}

\bibitem[\protect\citeauthoryear{{Miyatake} et~al.,}{{Miyatake}
  et~al.}{2019}]{Miyatake:2019}
{Miyatake} H.,  et~al., 2019, \mn@doi [\apj] {10.3847/1538-4357/ab0af0}, \href
  {http://adsabs.harvard.edu/abs/2019ApJ...875...63M} {875, 63}

\bibitem[\protect\citeauthoryear{Miyazaki et~al.}{Miyazaki
  et~al.}{2015}]{Miyazaki:2015}
Miyazaki S.,  et~al., 2015, \mn@doi [Astrophys. J.]
  {10.1088/0004-637X/807/1/22}, 807, 22

\bibitem[\protect\citeauthoryear{{Miyazaki} et~al.,}{{Miyazaki}
  et~al.}{2018}]{Miyazaki:2018}
{Miyazaki} S.,  et~al., 2018, \mn@doi [\pasj] {10.1093/pasj/psx063}, \href
  {http://adsabs.harvard.edu/abs/2018PASJ...70S...1M} {70, S1}

\bibitem[\protect\citeauthoryear{{Nishimichi} et~al.,}{{Nishimichi}
  et~al.}{2018}]{Nishimichi:2018}
{Nishimichi} T.,  et~al., 2018, arXiv e-prints, \href
  {http://adsabs.harvard.edu/abs/2018arXiv181109504N} {p. 1811.09504}

\bibitem[\protect\citeauthoryear{{Planck Collaboration} et~al.,}{{Planck
  Collaboration} et~al.}{2014a}]{Planck2013I}
{Planck Collaboration} et~al., 2014a, \mn@doi [\aap]
  {10.1051/0004-6361/201321529}, \href
  {http://adsabs.harvard.edu/abs/2014A%26A...571A...1P} {571, A1}

\bibitem[\protect\citeauthoryear{{Planck Collaboration} et~al.,}{{Planck
  Collaboration} et~al.}{2014b}]{Planck2013XVII}
{Planck Collaboration} et~al., 2014b, \mn@doi [\aap]
  {10.1051/0004-6361/201321543}, \href
  {https://ui.adsabs.harvard.edu/abs/2014A&A...571A..17P} {571, A17}

\bibitem[\protect\citeauthoryear{{Planck Collaboration} et~al.,}{{Planck
  Collaboration} et~al.}{2016}]{Planck2015XIII}
{Planck Collaboration} et~al., 2016, \mn@doi [\aap]
  {10.1051/0004-6361/201525830}, \href
  {https://ui.adsabs.harvard.edu/abs/2016A&A...594A..13P} {594, A13}

\bibitem[\protect\citeauthoryear{{Reyes}, {Mandelbaum}, {Gunn}, {Nakajima},
  {Seljak}  \& {Hirata}}{{Reyes} et~al.}{2012}]{Reyes:2012}
{Reyes} R.,  {Mandelbaum} R.,  {Gunn} J.~E.,  {Nakajima} R.,  {Seljak} U.,
  {Hirata} C.~M.,  2012, \mn@doi [\mnras] {10.1111/j.1365-2966.2012.21472.x},
  \href {http://adsabs.harvard.edu/abs/2012MNRAS.425.2610R} {425, 2610}

\bibitem[\protect\citeauthoryear{{Rowe} et~al.,}{{Rowe}
  et~al.}{2015}]{Rowe:2015}
{Rowe} B.~T.~P.,  et~al., 2015, \mn@doi [Astronomy and Computing]
  {10.1016/j.ascom.2015.02.002}, \href
  {http://adsabs.harvard.edu/abs/2015A%26C....10..121R} {10, 121}

\bibitem[\protect\citeauthoryear{{Shirasaki}, {Takada}, {Miyatake},
  {Takahashi}, {Hamana}, {Nishimichi}  \& {Murata}}{{Shirasaki}
  et~al.}{2017}]{Shirasaki:2017}
{Shirasaki} M.,  {Takada} M.,  {Miyatake} H.,  {Takahashi} R.,  {Hamana} T.,
  {Nishimichi} T.,   {Murata} R.,  2017, \mn@doi [\mnras]
  {10.1093/mnras/stx1477}, \href
  {https://ui.adsabs.harvard.edu/abs/2017MNRAS.470.3476S} {470, 3476}

\bibitem[\protect\citeauthoryear{{Shirasaki}, {Hamana}, {Takada}, {Takahashi}
  \& {Miyatake}}{{Shirasaki} et~al.}{2019}]{Shirasaki:2019}
{Shirasaki} M.,  {Hamana} T.,  {Takada} M.,  {Takahashi} R.,   {Miyatake} H.,
  2019, \mn@doi [\mnras] {10.1093/mnras/stz791}, \href
  {http://adsabs.harvard.edu/abs/2019MNRAS.486...52S} {486, 52}

\bibitem[\protect\citeauthoryear{{Simon} et~al.,}{{Simon}
  et~al.}{2013}]{Simon:2013}
{Simon} P.,  et~al., 2013, \mn@doi [\mnras] {10.1093/mnras/stt069}, \href
  {https://ui.adsabs.harvard.edu/abs/2013MNRAS.430.2476S} {430, 2476}

\bibitem[\protect\citeauthoryear{{Spergel} et~al.,}{{Spergel}
  et~al.}{2015}]{Spergel:2015}
{Spergel} D.,  et~al., 2015, arXiv e-prints, \href
  {http://adsabs.harvard.edu/abs/2015arXiv150303757S} {p. 1503.03757}

\bibitem[\protect\citeauthoryear{{Sunyaev} \& {Zeldovich}}{{Sunyaev} \&
  {Zeldovich}}{1969}]{Sunyaev:1969}
{Sunyaev} R.~A.,  {Zeldovich} Y.~B.,  1969, \mn@doi [\nat] {10.1038/223721a0},
  \href {https://ui.adsabs.harvard.edu/abs/1969Natur.223..721S} {223, 721}

\bibitem[\protect\citeauthoryear{{Sunyaev} \& {Zeldovich}}{{Sunyaev} \&
  {Zeldovich}}{1972}]{Sunyaev:1972}
{Sunyaev} R.~A.,  {Zeldovich} Y.~B.,  1972, Comments on Astrophysics and Space
  Physics, \href {https://ui.adsabs.harvard.edu/abs/1972CoASP...4..173S} {4,
  173}

\bibitem[\protect\citeauthoryear{{Takada} et~al.,}{{Takada}
  et~al.}{2014}]{Takada:2014}
{Takada} M.,  et~al., 2014, \mn@doi [\pasj] {10.1093/pasj/pst019}, \href
  {http://adsabs.harvard.edu/abs/2014PASJ...66R...1T} {66, R1}

\bibitem[\protect\citeauthoryear{{Takahashi}, {Hamana}, {Shirasaki},
  {Namikawa}, {Nishimichi}, {Osato}  \& {Shiroyama}}{{Takahashi}
  et~al.}{2017}]{Takahashi:2017}
{Takahashi} R.,  {Hamana} T.,  {Shirasaki} M.,  {Namikawa} T.,  {Nishimichi}
  T.,  {Osato} K.,   {Shiroyama} K.,  2017, \mn@doi [\apj]
  {10.3847/1538-4357/aa943d}, \href
  {http://adsabs.harvard.edu/abs/2017ApJ...850...24T} {850, 24}

\bibitem[\protect\citeauthoryear{{Tanaka} et~al.,}{{Tanaka}
  et~al.}{2018}]{Tanaka:2018}
{Tanaka} M.,  et~al., 2018, \mn@doi [\pasj] {10.1093/pasj/psx077}, \href
  {http://adsabs.harvard.edu/abs/2018PASJ...70S...9T} {70, S9}

\bibitem[\protect\citeauthoryear{{York} et~al.,}{{York}
  et~al.}{2000}]{York:2000}
{York} D.~G.,  et~al., 2000, \mn@doi [\aj] {10.1086/301513}, \href
  {http://adsabs.harvard.edu/abs/2000AJ....120.1579Y} {120, 1579}

\bibitem[\protect\citeauthoryear{{Zhang}, {Dietrich}, {McKay}, {Sheldon}  \&
  {Nguyen}}{{Zhang} et~al.}{2013}]{Zhang:2013}
{Zhang} Y.,  {Dietrich} J.~P.,  {McKay} T.~A.,  {Sheldon} E.~S.,   {Nguyen} A.
  T.~Q.,  2013, \mn@doi [\apj] {10.1088/0004-637X/773/2/115}, \href
  {https://ui.adsabs.harvard.edu/abs/2013ApJ...773..115Z} {773, 115}

\bibitem[\protect\citeauthoryear{{Zheng} et~al.,}{{Zheng}
  et~al.}{2005}]{Zheng:2005}
{Zheng} Z.,  et~al., 2005, \mn@doi [\apj] {10.1086/466510}, \href
  {http://adsabs.harvard.edu/abs/2005ApJ...633..791Z} {633, 791}

\bibitem[\protect\citeauthoryear{{de Graaff}, {Cai}, {Heymans}  \&
  {Peacock}}{{de Graaff} et~al.}{2019}]{deGraaff:2019}
{de Graaff} A.,  {Cai} Y.-C.,  {Heymans} C.,   {Peacock} J.~A.,  2019, \mn@doi
  [\aap] {10.1051/0004-6361/201935159}, \href
  {http://adsabs.harvard.edu/abs/2019A%26A...624A..48D} {624, A48}

\bibitem[\protect\citeauthoryear{{de Jong}, {Verdoes Kleijn}, {Kuijken}  \&
  {Valentijn}}{{de Jong} et~al.}{2013}]{KiDS:2013}
{de Jong} J.~T.~A.,  {Verdoes Kleijn} G.~A.,  {Kuijken} K.~H.,   {Valentijn}
  E.~A.,  2013, \mn@doi [Experimental Astronomy] {10.1007/s10686-012-9306-1},
  \href {http://adsabs.harvard.edu/abs/2013ExA....35...25D} {35, 25}

\makeatother
\end{thebibliography}
